\documentclass[fleqn,final,3p,times,twocolumn,authoryear]{elsarticle}

\usepackage{amssymb}
\usepackage{color}
\usepackage{lineno}

\definecolor{revision}{rgb}{0,0,0}
\definecolor{comment}{rgb}{1,0,0}

\journal{arXiv}

\begin{document}

\begin{frontmatter}



\title{\textit{In vivo} evaluation of wearable head impact sensors}


\author{Lyndia C. Wu, Vaibhav Nangia, Kevin Bui, Bradley Hammoor, Mehmet Kurt, Fidel Hernandez, Calvin Kuo, David B. Camarillo}

\address{Stanford University, Stanford, CA, 94305}

\begin{abstract}
{\color{revision}Inertial sensors are commonly used to measure human head motion.} Some sensors have been tested with dummy or cadaver experiments with mixed results, and methods to evaluate sensors \textit{in vivo} are lacking. Here we present {\color{revision}an \textit{in vivo}} method using high speed video to test teeth-mounted (mouthguard), soft tissue-mounted (skin patch), and headgear-mounted (skull cap) sensors during {\color{revision}6-13g}  sagittal soccer head impacts. {\color{revision}Sensor coupling to the skull} was quantified by displacement from an ear-canal reference. Mouthguard displacements were within video measurement error ($<$1mm), while the skin patch and skull cap displaced up to 4mm and 13mm from the ear-canal reference, respectively. We used the mouthguard, {\color{revision}which had the least displacement from skull }, as the reference to assess 6-degree-of-freedom skin patch and skull cap measurements. Linear and rotational acceleration magnitudes were over-predicted by both the skin patch ({\color{revision}with 120\% NRMS error for $a_{mag}$, 290\% for $\alpha_{mag}$}) and the skull cap ({\color{revision}320\% NRMS error for $a_{mag}$, 500\% for $\alpha_{mag}$}). Such over-predictions were largely due to out-of-plane motion. To model sensor error, we found that in-plane skin patch acceleration peaks in the anterior-posterior direction could be modeled by an underdamped viscoelastic system. In summary, the mouthguard showed {\color{revision}tighter skull coupling than the other sensor mounting approaches}. Furthermore, the \textit{in vivo} methods presented are valuable for investigating skull acceleration sensor technologies.

\end{abstract}

\begin{keyword}
instrumented mouthguard \sep instrumented skin patch \sep instrumented skull cap \sep high speed video \sep soft tissue modeling 



\end{keyword}

\end{frontmatter}



\section{Introduction} Traumatic brain injury biomechanics can be studied in human subjects using wearable
head impact sensors that measure skull accelerations. The availability of low-power, low-cost MEMS
accelerometers and gyroscopes has spawned a flurry of head impact sensing
approaches both in research and for consumer use.  A helmet-mounted sensor system, the head
impact telemetry system (HITS), {\color{revision}is an example of a widely used sensing option \citep{Duma2005,Rowson2011e,Rowson2011j}.
However, factors such as helmet fit and padding type may affect sensor coupling
to a human head, and in turn cause measurement errors
\citep{Higgins2007d,Beckwith2012h,Jadischke2013}. More recently, industry and
academic labs have developed alternative sensors with other form factors and
mounting locations such as the teeth, ear-canal, skin, and various types of headgear. For these devices, factors including fit, adhesion, soft-tissue
elasticity, and hair/scalp properties may affect sensor skull coupling and
measurement accuracy. 

Instrumented bite blocks have been used \textit{in vivo} as reference sensors
\citep{Funk2009,Knox2004}. A similar approach is to instrument a
mouthguard, which is practical for field use in contact sports.  Instrumented
mouthguards have been evaluated \textit{in vitro} with a clamped-jaw dummy as a
reference \citep{Camarillo2013a}. One error source was introduced when the
sensors were placed in a protruding tab on the mouthguard, which exhibited a
mechanical resonance. The resonance led to errors in peak acceleration
measurements, but RMS acceleration errors were still within 10\%.
\citet{Bartsch2014} also demonstrated accuracy using a dummy head that does not
have a lower jaw to clamp the mouthguard. 

Skin patch and skull cap sensors are also used in research and are
becoming commercially available. Skull coupling of either approach has not
been evaluated in literature. Previous studies with skin-mounted sensors and
optical markers at other locations on the body (e.g. knee joint) report
artifacts due to skin dynamics
\citep{Reinschmidt1997,Lucchetti1998,shultz2011quantifying}. Differentiation of position and orientation measurements can amplify acceleration errors from such soft tissue artifacts. Therefore, some researchers
have modeled soft-tissue dynamics to correct for measurement errors
\citep{Trujillo1990,Kim1993}. It is likely that soft tissue or textile dynamics
affects sensor performance, but the effects have not been quantified for head
impacts.

{\color{revision}Sensors are commonly evaluated through \textit{in vitro} (anthropomorphic test device) and/or \textit{ex vivo}
(postmortem human subject) methods.  Both methods have the advantage of using high fidelity reference sensors rigidly attached to the skull.} methods test sensor accuracy without confounding
factors such as soft tissue motion, and are valuable for verifying sensor
hardware selection and programming. \textit{Ex vivo} methods introduce
additional biofidelic factors including skull and tissue dynamics, but the
effects of postmortem changes in tissue properties are unknown, and the lack of
muscle forces may also affect head dynamics. Thus \textit{in vivo} sensor evaluation may help to account for these factors. But there is a lack of \textit{in vivo} methods, since 
we can neither screw reference
sensors directly to the skull, nor use dangerous impact conditions in human subject studies. 

{\color{revision}Thus our
objective is to develop a non-invasive \textit{in vivo} method to evaluate head impact sensing approaches. Using this method, we will test three types of approaches: sensors fit to hard tissue (teeth), adhered to soft tissue (skin), or mounted on headgear that fits the head. Corresponding to these sensor types, we assessed skull coupling of instrumented
mouthguard, skin patch, and skull cap sensors in a human subject during mild
soccer head impacts using high speed video tracking (Fig.  \ref{fig:methodology_illustration}, Methods Sec 2.3, Results Sec 3.1). Then we used the sensor with the least amount of skull displacement, which is the mouthguard, as the reference sensor to compare full 6-degree-of-freedom (6DOF) sensor measurements (Methods Sec 2.4, Results Sec 3.2). To account for sensor error from soft-tissue motion, we modeled sensor-skull dynamics using a simple viscoelastic model (Methods Sec 2.5, Results Sec 3.3).}

\section{Methods} 

\subsection{Human Subject Experiment Setup}
A 26 year-old male human subject underwent soccer
head impacts with clenched teeth, at initial ball speed of 7m/s
(Fig. \ref{fig:methodology_illustration}A), which is the average header speed in
youth soccer, and higher than that (5.7m/s) in adults \citep{Shewchenko2005a}. A
ball launcher (Sports Tutor, Burbank, CA) helped to simulate a kicked ball, and
the ball was inflated to approximately 8-9psi. Human subject protocols in this study have been approved by the Stanford
Institutional Review Board (IRB No. 26620), and informed consent was obtained
from the subject. 

\subsection{Instrumentation}
The volunteer wore a custom-fit mouthguard
\citep{Wu2014a}, a skin patch adhered to skin on the mastoid process (xPatch Gen2, X2Biosystems, Inc.), {\color{revision}and an elastic
skull cap (Reebok).} The mouthguard had approximately 4mm average thickness and 7mm
height above the gum line.  Electronics were placed inside the mouth to avoid
tab resonance, while video markers were fixed on a light-weight protruding tab.  The
volunteer's head circumference measured 60.3cm, and wore a size large elastic
skull cap. The same electronics from the
mouthguard were placed on a soft cardboard in the lateral insert of the skull cap. To enable comparison of measurements, devices were set to a low triggering threshold (4 - 6g) to ensure all
impacts were recorded on each device.  Sensor signals were synchronized through video{\color{revision} by aligning sensor-measured kinematics with video-derived sensor kinematics for each individual sensor. To derive sensor kinematics from video, we resolved the position and orientation of each sensor using a previously-described method of tracking fiducial markers in stereo video \citet{Hernandez2015}.} During synchronization, we found that
skin patch {\color{revision}motion} can lag behind the mouthguard by as much as 15 ms. So we
recorded 30ms of pre-trigger data and 70ms post-trigger.

{\color{revision}In addition to these three sensors, the subject wore a deeply-inserted earplug as a skull reference point. Previous research confirmed
skull coupling of deeply-inserted custom-fit silicone earplugs using reference sensors
screwed onto the skull of a postmortem human subject \citep{Salzar2014,Christopher2013}. An
expandable foam earplug was inserted approximately 20mm into the ear canal,
similar to the depth at which an ear sensor in the postmortem experiment was mounted. The low mass of a foam earplug (0.2g, compared to 5g for a typical custom-formed silicone earplug) also minimizes inertial
effects and improves coupling to the ear canal. This skull reference was used to assess skull coupling of the three sensors.}

\subsection{Video analysis of skull coupling}   

We took high speed stereo video at 1000 frames per second and 1920$\times$1200 resolution (0.3mm/pixel at distance of head), using two Phantom
Miro LC320 cameras (Vision Research, Wayne, NJ), to track motion of the
sensors (Fig. \ref{fig:methodology_illustration}B).  Fiducial markers with 1.5mm
square grids were fixed onto each {\color{revision}sensor} and the earplug to track the change in distance between
grid centroids. {\color{revision}The mouthguard, skin patch, and skull cap sensors each had 26-30 trackable points, and the earplug had 1-3 trackable points.} The two cameras were
positioned such that 1) the tracking grid on the deeply inserted earplug had at
least 1 trackable point and 2) all three sensors were visible throughout the head impact. Due to these constraints, there
was a triangulation angle of 7.4 degrees between the cameras. Using the Camera
Calibration Toolbox for Matlab \citep{Bouguet2013, zhang1999flexible, heikkila1997four}, we performed stereo
calibration to enable 3D position tracking in the head motion space. 

To verify our video method, we tracked the 4 corners of a 20cm$\times$20cm calibration grid during 6DOF motion through
4000 frames (4 seconds). {\color{revision}The calibration grid moved through the same region of space as head motion, at the same distance away from the cameras as the subject.}{\color{comment}(R3-4)} We assessed both planar and depth measurements of distance, by
comparing stereo video measurements with ground truth grid distances in each
frame. In addition, we derived sagittal kinematics of each sensor from video
measurements, to cross validate with sensor measurements. {\color{revision}Position time histories of points fixed on the grid were combined to determine the least-squares rotation matrix describing the orientation of the body-fixed frame of each sensor in the camera-fixed frame \citep{Hernandez2015}. The time derivative of this rotation matrix was related to the rotational velocity of the body-fixed frame in the camera-fixed frame. We used this rotational velocity to take the time derivative of linear position and velocity to determine velocity and acceleration, respectively, in the moving body-fixed frame.} Due to
near-parallel arrangement of cameras, out-of-plane measurements are expected to
have larger errors, and we compared 3DOF sagittal kinematics instead of full
6DOF kinematics when validating our video method.

\subsection{Comparison of 6-DOF sensor kinematics} Video measurements of the ear-canal reference only allowed for 1-3 trackable
points, so 6DOF kinematics could not be computed. To evaluate 6DOF
measurement differences, we selected a reference device with the least
relative motion to the ear-canal - the mouthguard sensor. 

We transformed kinematic measurements of the skin patch and skull cap sensors to
{\color{revision}an estimated center-of-gravity (CG) location}, for trials where mouthguard and skull were found to
have the lowest relative displacement (see Results Sec. 3.1). {\color{revision}CG location was estimated based on a 50th percentile male human head model. We first estimated the projection vector from mouthguard to CG by using the upper dentition as an anatomical landmark on the model. Then we measured the projection vectors of the skin patch and skull cap sensors to the mouthguard location on our human subject. Using stereo video of the subject standing still without head motion, we derived relative position and
orientation of the 3 devices and confirmed with
physical measurements. The resultant projection vectors for the skin patch and skull cap sensors were calculated by summing their projections to mouthguard and the projection from mouthguard to CG. Using this method, we ensure that all three devices can be projected to the same point in space for comparison, even though the model-estimated CG location is expected to have some error.} 

Linear accelerations {\color{revision}were projected to the CG} location for comparison using the following
rigid body vector relationship:

\begin{equation}
\vec{{a}}_{CG} = \vec{{a}}_{s} + \vec{{\alpha}} \times \vec{{r}}_{s} + \vec{{\omega}} \times (\vec{{\omega}} \times \vec{{r}}_{s}) \\
\label{eq:1}
\end{equation}

\noindent {\color{revision}where $a_{CG}$ is head linear acceleration at CG, $a_s$ is head linear acceleration at each sensor location,
$\alpha$ is head angular acceleration measured by the sensor, $\omega$
is head angular velocity measured by the sensor, and $r_s$ is the
vector position of CG location from the sensor location.  }

Transformed skin patch and skull cap sensor data were compared with mouthguard reference data in the anterior-posterior (AP), left-right (LR),
inferior-superior (IS) directions for linear acceleration, and the coronal,
sagittal, and horizontal planes for rotational acceleration. Quantities reported
include vector magnitudes and individual axis differences from all 6DOF linear
acceleration, angular velocity, and angular acceleration. {\color{revision}We performed linear regression analysis of peak kinematic values.} We also reported the
average deviation in peak values from the mouthguard reference, and compared the
directions of head acceleration.  In addition to peak values, we assessed the
agreement of time traces by computing root-mean-square (RMS) difference and
normalized root-mean-square (NRMS) difference for 25 samples around peak
measurements \citep{Camarillo2013a}. {\color{revision}To better understand the sources of sensor errors prior to projection to CG, we also computed the RMS and NRMS differences at the sensor location. That is, we compared skin patch/skull cap signals at the skin patch/skull cap location with mouthguard reference signals projected to the skin patch/skull cap location.}

\subsection{Modeling of sensor dynamics} 
\subsubsection{Model description}Since the soccer headers were frontal hits and head motion was mainly
anterior-posterior, we modeled the skin patch and skull cap translation in the
anterior-posterior direction as a second order linear system with base (skull)
excitation (Fig. \ref{fig:dynamic_model}A, Equation \ref{eq:2}), where
\begin{itemize}
	\item[] $m_{sensor}$ - combined mass of the sensor and soft tissue (kg)
	\item[] $K_{t}$ - effective linear spring constant of sensor-skull mounting (N/m)
	\item[] $C_{t}$ - effective linear damping constant of sensor-skull mounting (N-s/m)
	\item[] $d_{sensor}$ - absolute displacement of the sensor (m)
	\item[] $d_{skull}$ - absolute displacement of the skull (m)
\end{itemize}

\noindent
\begin{equation}
{m}_{sensor} \ddot{d}_{sensor} = - {K}_{t} ({d}_{sensor} - {d}_{skull}) - {C}_{t} (\dot{d}_{sensor} - \dot{d}_{skull}) \\\label{eq:2}
\end{equation}

We also modeled sagittal rotation of the head using a similar second order linear system (Fig. \ref{fig:dynamic_model}B, Equation \ref{eq:3}).

\noindent
\begin{equation}
{I}_{sensor} \ddot{\theta}_{sensor} = - {K}_{r} ({\theta}_{sensor} - {\theta}_{skull}) - {C}_{r} (\dot{\theta}_{sensor} - \dot{\theta}_{skull}) \\
\label{eq:3}
\end{equation}

\subsubsection{Model evaluation}We used sensor data from soccer impacts to find the input-output relationship
defining the skull-tissue-sensor system. Only trials where mouthguard-skull
coupling was {\color{revision}best} ($<$0.5mm) were modeled. We fit model parameters
using mouthguard measurements as skull input, and skin patch/skull cap
measurements as sensor output. 

{\color{revision}Sensor modeling was done by fitting spring, damper, and mass parameters to the input skull kinematics and output sensor kinematics. This system takes prescribed skull motion (obtained by transforming the mouthguard kinematic signal to the sensor location) as an input and does a forward dynamics simulation using the Matlab ode45 integrator to solve for the resulting sensor kinematics. The initial state for the system is set such that there is no relative motion between the skull and sensor and the spring is at its rest length. We used the Matlab function fmincon to find a set of spring stiffness, damping coefficient, and mass parameters that minimize the root mean squared error between the output model sensor kinematics and the measured sensor kinematics. The search space for these parameters are bounded to constrain the system to realistic values and the initial guess is set as the midway point between bounds.  Considering that the
sensors moved along with packaging and underlying soft tissue, a loose mass
bound of 1 to 50 grams was placed on the optimization to determine the mass of the
system. Each impact is separately fitted in order to account for variabilities in impact conditions (location and force). The NRMS of the fit was used to assess the model. }

In order to check the validity of the linearity assumption in our model, we
examined the input-output relationship in the frequency domain by plotting the
experimental and theoretical frequency response functions (FRFs). Experimental
FRFs were calculated by taking the ratios of the Fast Fourier Transform (FFT)
amplitudes of the skull input and skin patch output. Furthermore, the analytical frequency response of the
base-excitation model depicted in Fig. \ref{fig:dynamic_model}A was compared against the experimental
FRFs.

\section{Results}
\subsection{Skull coupling from video}
We verified stereo video tracking to have $<$1mm error in the sagittal plane (Fig.
\ref{fig:video_results}A). When the calibration grid was displaced
or rotated in a plane perpendicular to the camera axis (sagittal), stereo
triangulation estimated {\color{revision}distances between points on the calibration grid} with $<$1mm error. When depth
measurement was involved, such as during out-of-plane rotation (i.e. grid was tilted from camera plane), errors were slightly
larger but still within 2mm. Overall, more than 90\% of the errors were within 0.5mm. As an additional verification step, we compared
video-derived sagittal kinematics  {\color{revision}of each sensor with those measured by the accelerometer and gyroscope in each sensor. These kinematics matched with
$<$30\% NRMS difference}, even for linear acceleration
double-differentiated from position tracking (Fig. \ref{fig:video_results}C,
Table \ref{table:video_validation_results}).

For the mouthguard, all head impact trials (n=16) showed relative displacements from
the earplug of $<$1mm (Fig. \ref{fig:coupling_results}, $\mu$=0.5mm,
$\sigma$=0.2mm), within video measurement error. In
addition, 10 of the 16 trials showed relative mouthguard displacements of within
0.5mm. In contrast, the skin patch sensor
displaced by 2-4mm ($\mu$=3mm, $\sigma$=0.7mm) at the moment of head impact; the skull cap sensor
displaced by 2-13mm ($\mu$=5mm, $\sigma$=3mm).

\subsection{6-DOF sensor kinematics}
Using the mouthguard as a reference, head acceleration peak magnitudes for the
10 trials averaged {\color{revision}9.3$\pm$2g} in linear acceleration and 750$\pm$300$rad/s^2$
for rotational acceleration. Skin patch estimation of head linear and angular
acceleration peak values were over-predicted by {\color{revision}15$\pm$7g} and
2500$\pm$1200$rad/s^2$ on average, respectively; {\color{revision}the skull cap values were over-predicted by} {\color{revision}50$\pm$31g} and 4300$\pm$2700$rad/s^2$ (Table
\ref{table:peak_difference_results}). {\color{revision}These over-predictions are the average differences between the sensor measurements and the reference (mouthguard) measurements} Fig.  \ref{fig:lin_regression_of_peaks}A
also shows the over-predictions in peak magnitudes, with patch/cap predictions
scattered above and away from the m=1 reference line. In addition, patch and cap peak vector magnitudes and all individual component peaks had large
variances, and did not correlate well
with the mouthguard reference {\color{revision}(Supplemental Fig. \ref{supfig:peak_scatter}). When y-intercept is forced to be 0 in the regression, the coefficient of determination is sometimes negative (Supplemental Table \ref{suptable:lin_reg_results}), indicating that the residual error in the fit is greater than the variance in the data.} To evaluate dynamic relationships among sensors,
Table \ref{table:peak_difference_results} reports the time lag/lead between the
mouthguard peak value and patch/cap peak values.  The skin patch linear
acceleration magnitude peak was the most consistent, occurring 15$\pm$3ms after
the mouthguard peak.

{\color{revision}The skin patch and skull cap also predict different directions of head motion compared with the mouthguard.} Fig.
\ref{fig:lin_regression_of_peaks}B and Table \ref{table:peak_difference_results}
show the differences in kinematic vectors at the moment of peak magnitude. The
mouthguard reference measured head motion to exhibit mostly planar motion with
{\color{revision}anterior-posterior (AP)} linear acceleration and sagittal rotation.  In contrast,
skin patch motion was mostly out-of-plane with left translation and horizontal rotation, and the
skull cap was not in a consistent direction. Breaking magnitudes down into per-component comparisons, Fig.
\ref{fig:accuracy_results_waveforms} shows sample 6DOF kinematic waveforms of a
representative impact with mostly anterior-posterior motion. We observe
over-predictions of kinematics in all axes, including out-of-plane (non-sagittal) directions. In fact, the highest
peaks for the skin patch all occur in out-of-plane axes: {\color{revision}left-right (LR)} linear acceleration
and horizontal rotation. The skin patch signals also show damped oscillatory
behavior. Table \ref{table:accuracy_RMS_results} shows RMS and NRMS errors of
the skin patch and skull cap in 6DOF. All magnitude errors were above 100\%. {\color{revision}At sensor location, the linear acceleration RMS and NRMS errors are lower than those at the CG (comparing Table \ref{table:accuracy_RMS_results} and Supplemental Table \ref{suptable:RMS_results_at_sensor}). Linear acceleration magnitude has 18\% NRMS error for the skin patch, and 60\% NRMS error for the skull cap, compared to 120\% and 320\% when projected to CG. Angular velocity and angular acceleration are independent of the location on the head, and have the same high errors at both sensor location and CG. }.

\subsection{Sensor modeling}
Identified model parameters are detailed in Table \ref{table:model_parameters},
including the NRMS errors of the fit. The statistics are generated from: all
trials for skin patch AP translation; 7 of 10 trials for skin patch sagittal
rotation; and 8 of 10 trials for cap AP translation. {\color{revision}Since this is a very simplified model, we did not expect all trials to fit to the model. In order to report meaningful parameter estimates, we omitted trials where the model could not fit to the data.} Some trials were omitted
for patch sagittal rotation since the input mouthguard signal lagged behind the
patch measurement. In such cases, the mouthguard signal could not be used as
input for modeling. Instead, the input to the system likely came from other
coupled axes. Some trials were omitted for skull cap AP translation, since
reasonable estimates of model parameters could not be found due to sharp spikes
in the signal, which likely resulted from direct soccer ball impact. For all
trials of skull cap sagittal rotation, the mouthguard signal consistently lagged
behind the skull cap signal, and the model failed to fit. Fig.
\ref{fig:model_evaluation_waveforms} shows sample traces comparing the model fit
and measured signals.  Overall, model fits were the most consistent for skin
patch in AP translation. The average estimated mass of the skin patch system
(8.5g) is greater than mass of the sensor itself (5g), showing that some
underlying tissue mass was also activated during the impacts.

For skin patch AP translation, where the model could be fit to all 10 trials
with relatively low variance in model parameters, we verified the linear system
assumption and examined the frequency response around the resonant frequency of
the model. Fig. \ref{fig:freq_response_fn}A and B show the FFT amplitudes for
the mouthguard (skull) input and skin patch (sensor) output, respectively. The
FRFs, which are the ratio of output FFT amplitude to the input FFT amplitude,
are shown for all trials in Fig. \ref{fig:freq_response_fn}C. We compared these
with a theoretical FRF (red line), which is the frequency response of the
transfer function for the model in Fig.  \ref{fig:dynamic_model} with the
average model parameters (Table \ref{table:model_parameters}). As shown, 9 of
the 10 trials had FRFs with amplitudes and peak frequencies within a $\pm$1
standard deviation confidence interval, which validates the linearity
assumption, since a linear system exhibits a consistent FRF
\citep{ewins2000modal}.

\section{Discussion}
In this study, we developed an \textit{in vivo} method to quantify skull coupling of teeth-mounted (mouthguard), soft tissue-mounted (skin patch), and headgear-mounted (skull cap) sensors. All mouthguard-skull displacements were within video measurement
error. This was an expected outcome for mild impacts with
clenched teeth, since the mouthguard is custom-formed to the teeth. {\color{revision}It does
confirm that \textit{in vivo} clenching force can hold the mouthguard in place for the soccer impact conditions tested.} On the field, other practical factors such as mandible motion and variations in
mouthguard fabrication may affect performance. {\color{revision}Unanticipated impacts with open jaw may also pose a problem for the mouthguard.} Thus field
mouthguard coupling remains to be tested. However, in anticipated impacts, the
teeth are likely clenched, resulting in similar conditions as our tested
scenario.

Non-rigid skull coupling led to skull measurement errors in skin patch and skull
cap sensors. Both sensor
magnitudes showed over-predictions of peak {\color{revision}head CG} kinematics. Such over-predictions
resulted from measurements of significant out-of-plane motion, while head motion was
mostly sagittal. In fact, the skin patch and skull cap often measured
acceleration peaks in a different vector direction from the mouthguard reference
with $>$50 degrees deviation (Table
\ref{table:peak_difference_results}). With the over-predictions and differences in direction, raw data from these sensors likely cannot be directly used to predict or
study injury risks. Sensor errors need to be corrected via models, and/or
reduced by improving skull coupling. 

In the primary plane of motion (sagittal), a simple viscoelastic model
approximated single-degree-of-freedom skin-sensor dynamics, in agreement with
previous studies of lower-limb soft tissue dynamics
\citep{Trujillo1990,Kim1993}. As a feature of a linear dynamic system, the time
lag between skin patch translation and skull translation was consistently around
13ms (Table \ref{table:peak_difference_results}). {\color{revision}This time lag was due to viscoelasticity of the skull-skin-sensor system.} Skin patch motion in this axis had linear model fits with low variance in
parameters (Table \ref{table:model_parameters}). From the frequency response of
the system (Fig. \ref{fig:freq_response_fn}), we further confirm the underlying
linear dynamics of the system. For the soccer impact, the skull input excited a frequency range that
{\color{revision}includes the resonant frequency of the skull-skin-sensor system (20-30Hz), where the gain of the system is maximum}. However, if the head is driven at a different input
frequency (i.e. a different input duration) and/or at considerably higher
amplitudes, the peak gain may vary, or a different mode of the system may be
excited. Therefore, it would be an oversimplification to use a static gain term
to estimate head acceleration. 

Behavior of the skull cap sensor, especially in rotation, was less predictable
than the skin patch sensor, varying from impact to impact. The time lag between
cap and skull motion had large variance (Table \ref{table:model_parameters}),
indicating inconsistent behavior. This is likely due to direct impact of the
skull cap by the ball. Ball force and impact location were important factors for
the skull cap system not accounted for by the simple linear model. During our
experiments, the skull cap sometimes completely dislocated from the head
(results not included since skull cap was not in camera view). However, it is
possible that for other impact conditions, such as helmeted impacts, the skull
cap may have less relative motion from the skull.

{\color{revision} Although the modeling method shows promise in simulating soft tissue behavior, this may not be sufficient to correct for sensor error. Skin patch linear acceleration in the anterior-posterior direction had the lowest variance in the optimized model parameters. To estimate skull input from skin patch measurements, an inverse dynamics simulation can be performed using average model parameters. But this would only help to mitigate skin patch errors in 1DOF, and the variability in tissue response across different impact conditions and subjects can pose a challenge in developing a universal model. Also, referring to Supplemental Table \ref{suptable:RMS_results_at_sensor}, we show that the linear acceleration errors at head CG are much higher than those at the sensor locations. This indicates amplification of rotational velocity and acceleration errors when measurements are projected to CG. Thus  better models or design changes to mitigate rotational errors may help significantly improve sensor performance.}

The \textit{in vivo} methods in this study
have some limitations. First, only mild impacts were assessed in a single human subject. We tested
low-speed 7m/s impacts for protection of the human subject,
while field ball speeds could reach up to 17m/s \citep{Shewchenko2005a}. {\color{revision}Also, with one subject and one impact condition, the outcomes are likely subject to variability in soft tissue properties, skull cap fit, mouthguard fit, and impact location/severity.}  {\color{revision}The head acceleration levels (6-13g) are also low compared with injury-level accelerations on the field \citep{Hernandez2014}, for protection of human subjects.} Second, high speed stereo video tracking was limited by the
need to deeply insert the ear-canal reference for tight skull coupling. This led
to near-parallel arrangement of cameras and low number of trackable ear-canal
points. As a result, we could not derive 6DOF reference measurements from the skull reference. {\color{revision}Third, we made the assumption that the ear canal skull reference is rigidly attached to the skull. This limitation is difficult to eliminate in an \textit{in vivo} study. However, the fact that the mouthguard and the ear reference showed low relative motion gives confidence by cross validating these two mounting locations approximately 10cm apart on the head.} Fourth, mouthguard bite force was not controlled or measured
in the experiment. At the low acceleration levels in this experiment, we do
not expect bite force to significantly change results. However, at higher accelerations, bite force
may need to be quantified. Lastly, in practice, higher degree-of-freedom dynamic models
with more elements/parameters may be necessary to predict both in-plane and out-of-plane sensor errors.

In summary, we {\color{revision}have developed a} method to quantify skull coupling of wearable head impact
sensors \textit{in vivo}, and evaluated some common sensing approaches. The instrumented mouthguard was shown to have close skull coupling when clenched during mild soccer head impacts. The skin patch and skull cap devices had higher displacements from the skull. Raw data from sensors without
close skull coupling should be interpreted cautiously both in trauma research
and clinical assessment. To mitigate insufficient coupling, design modifications and modeling may help to
reconstruct skull motion.

\vspace{18pt}
\noindent \textbf{Conflict of interested statement} The authors have no personal or financial conflicts of interest related to this study.

\vspace{18pt}
\noindent \textbf{Acknowledgements}
We thank X2 Biosystems Inc. for supplying skin patch sensors. The study was supported by the National Institutes of Health (NIH) National Institute of Biomedical Imaging and Bioengineering (NIBIB) 3R21EB01761101S1,  Lucile Packard Foundation 38454, and the Stanford Child Health Research Institute Transdisciplinary Initiatives Program.

\section{Bibliography}
  \bibliographystyle{elsarticle-harv} 
  \bibliography{sensor_evaluation}


\clearpage
\begin{table*}[!ht]
\caption{RMS errors of video-tracked sagittal kinematics with respect to sensor measurements}
\includegraphics[width = \textwidth]{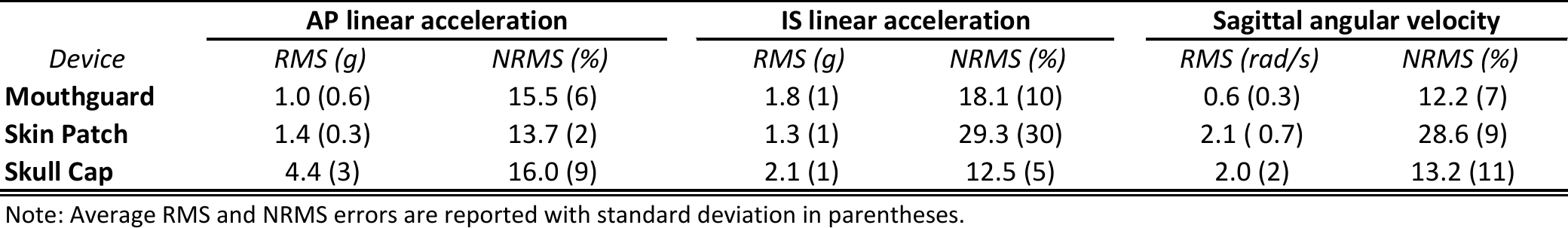}
\centering
\label{table:video_validation_results} 
\end{table*}

\clearpage
\begin{table*}[!ht]
\caption{Comparing skin patch and skull cap kinematic peak values with mouthguard reference}
\includegraphics[width = \textwidth]{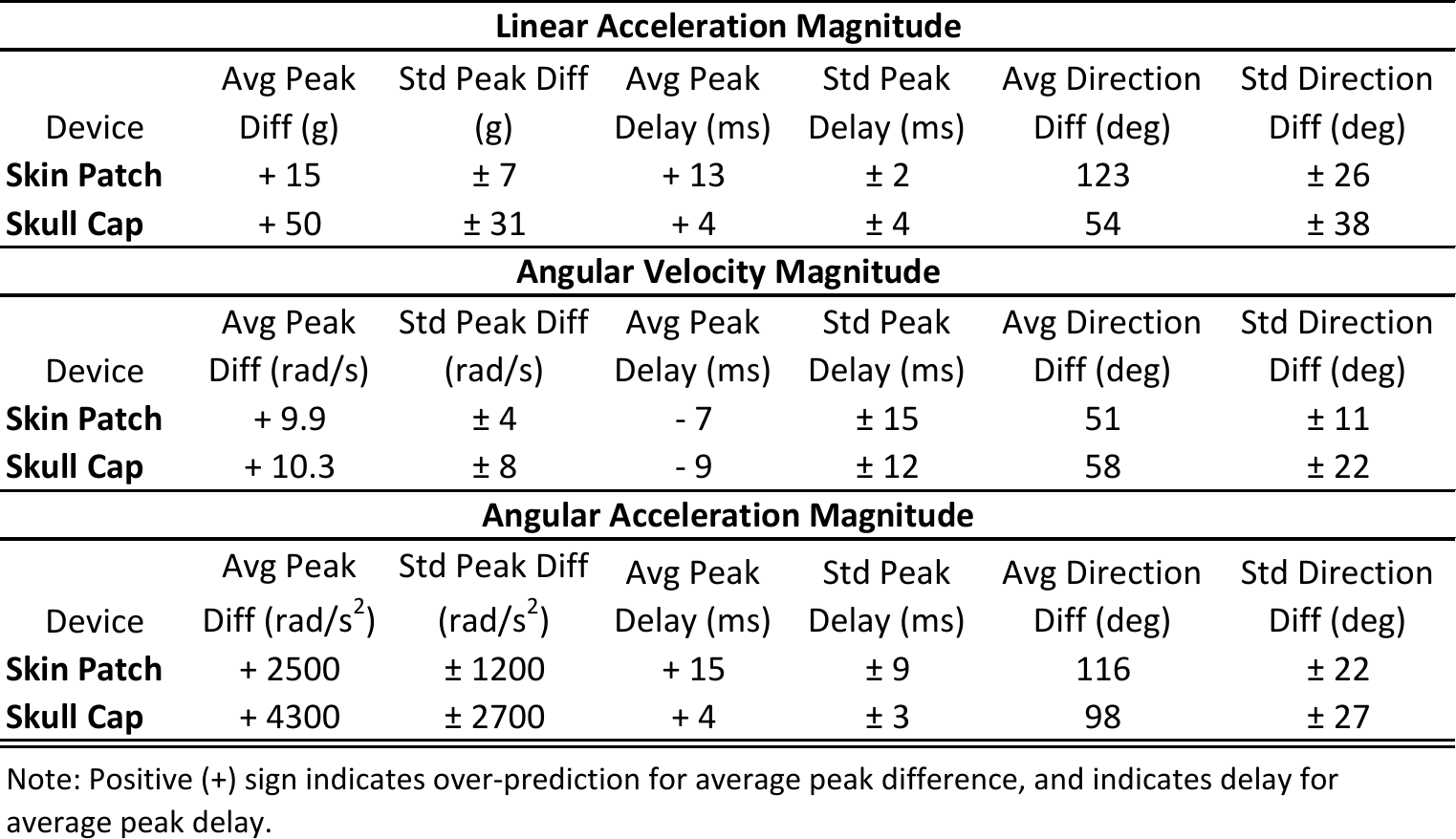}
\centering
\label{table:peak_difference_results} 
\end{table*}

\clearpage
\begin{table*}[!ht] 
\caption{Sensor RMS differences with respect to mouthguard time traces at head CG} 
\includegraphics[width = \textwidth]{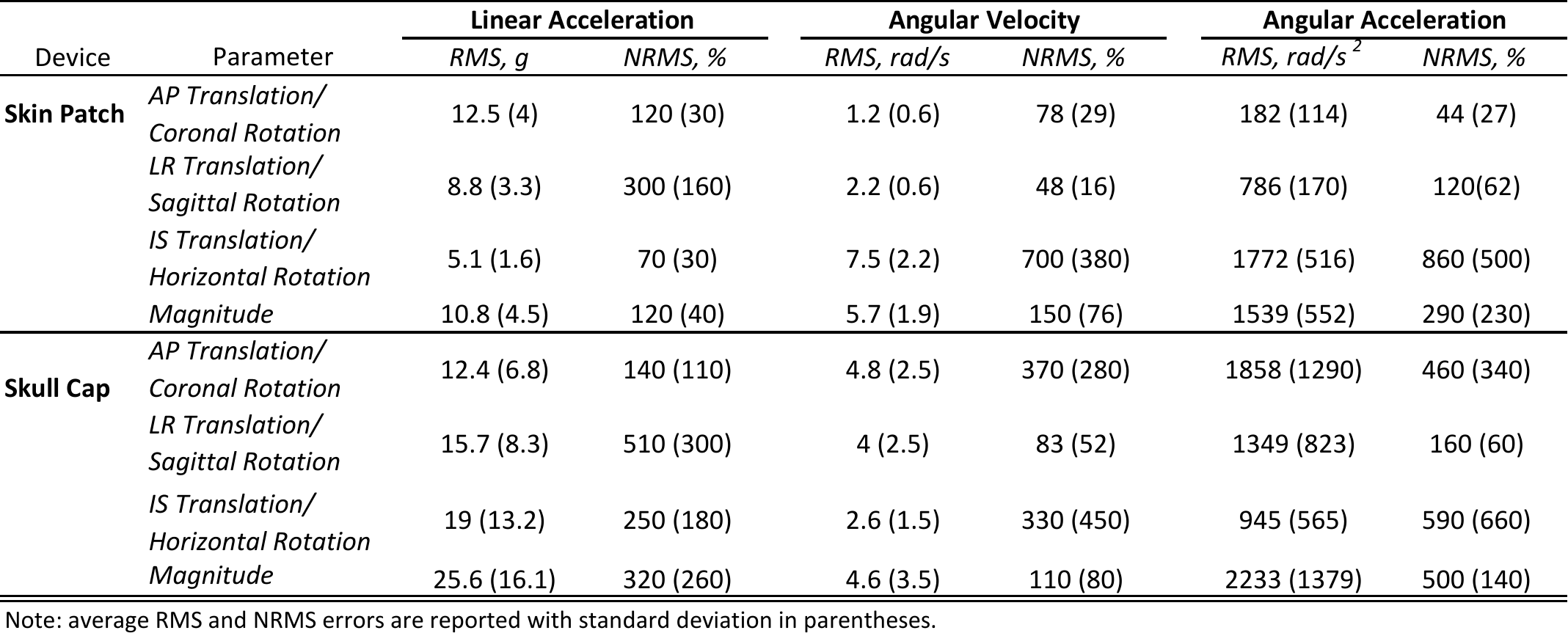}
\centering 
\label{table:accuracy_RMS_results}  
\end{table*}

\clearpage
\begin{table*}[ht] 
\caption{Model parameters}
\centering
\includegraphics[width = \textwidth]{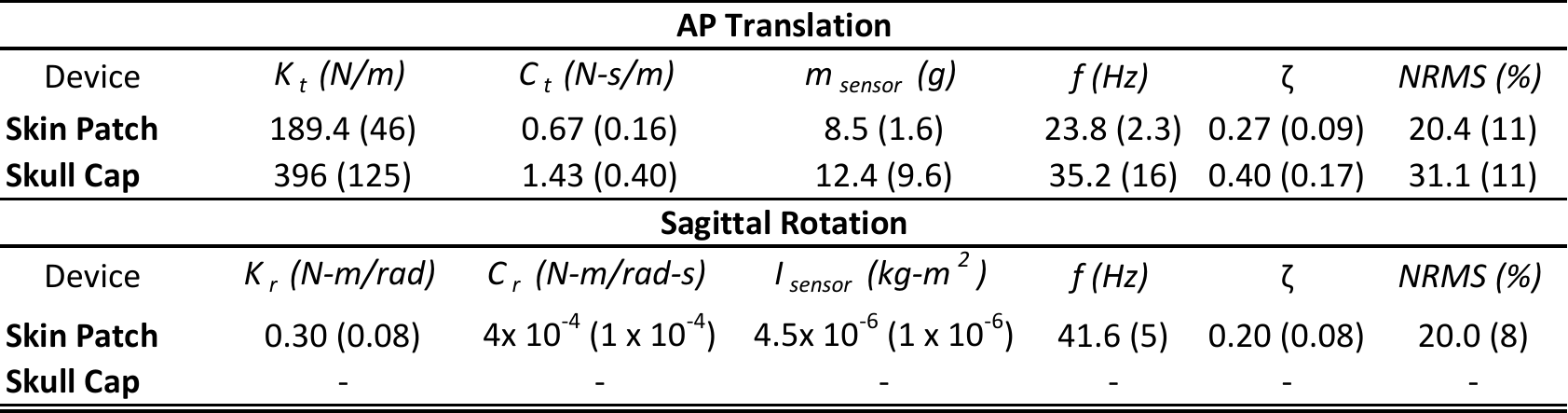}
\label{table:model_parameters} 
\end{table*}


\clearpage
\begin{figure}
\centering
\includegraphics[width = 0.8\textwidth]{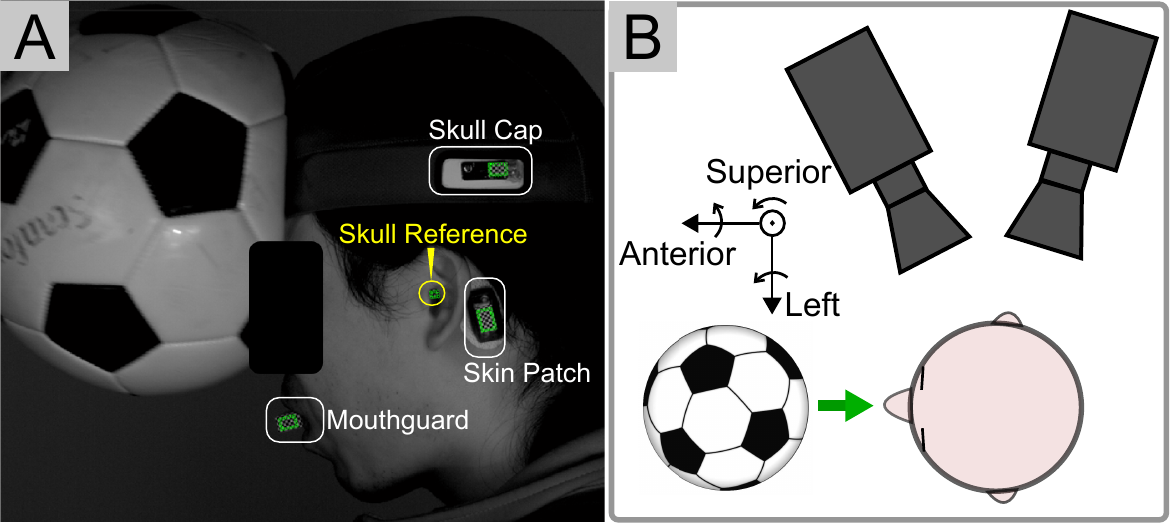}
\caption{
 \textit{In vivo} evaluation and comparison of instrumented mouthguard, skin patch, and skull cap. (A) A human subject underwent mild soccer head impacts, wearing all three sensors. Fiducial markers were mounted on the head, with one set on a deeply-inserted earplug (skull reference), and a set on each sensor. (B) Markers were tracked using high-speed stereo video to determine the relative motion between each sensor and the skull reference. }
\label{fig:methodology_illustration}
\end{figure}

\clearpage
\begin{figure}[h]
\includegraphics[width=0.8\textwidth]{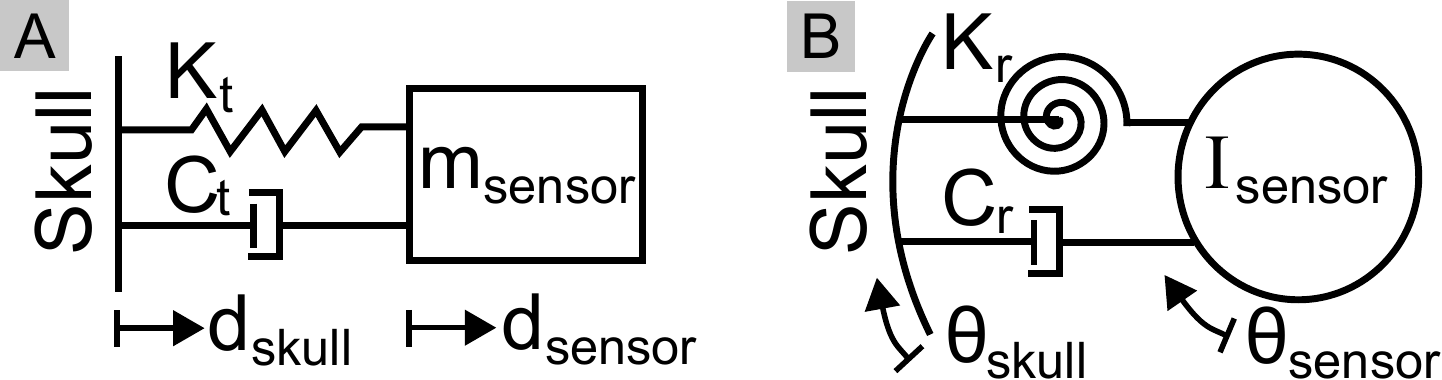}
\centering
\caption{Dynamic model description. (A) For anterior-posterior translation, we modeled each sensor-skull system as a second order linear system with a spring and a damper in parallel. These elements represent dynamics of the underlying tissue as well as the packaging and attachment of the sensors. (B) For the case of sagittal rotation, the elements are torsional springs and dampers.
}
\label{fig:dynamic_model}
\end{figure}

\clearpage
\begin{figure*}[t]
\centering
\includegraphics[width = \textwidth]{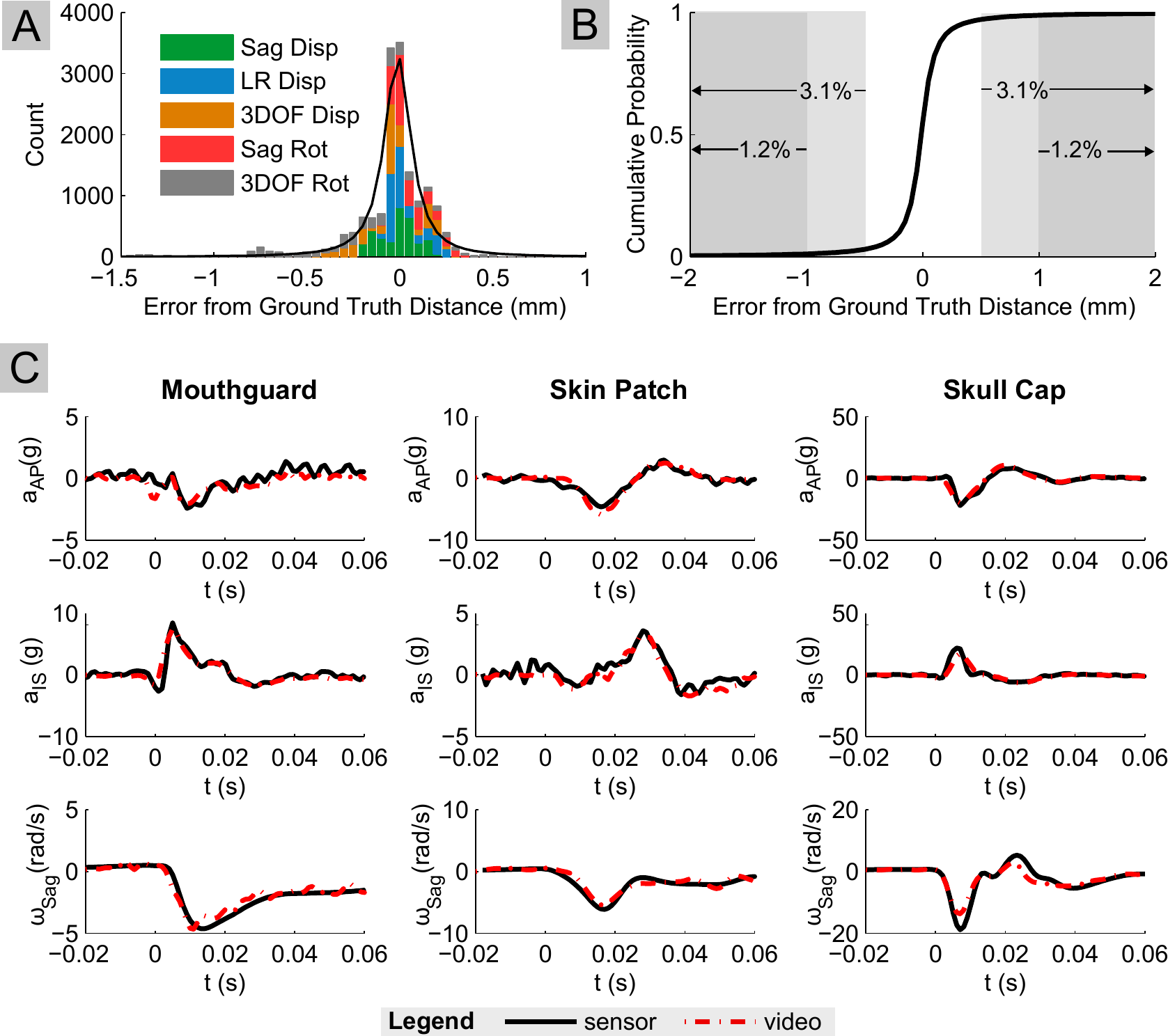}
\caption{
Video validation. (A) We calculated errors from video tracking using a 20cm x 20cm calibration grid moving through {\color{revision}the head motion region of space.} When the calibration grid displaced or rotated with a sagittal orientation (i.e. planar measurements), errors were always sub-millimeter. When depth measurement was involved, with the grid rotating in non-sagittal directions, errors were larger but still within 2mm. (B) We fit a t location-scale distribution to the error, and there is less than 2.5\% total probability of errors greater than 1mm. (C) We also verified that video-derived sagittal kinematics agree well with those measured by the sensors, which further confirms our video measurements.
 }
\label{fig:video_results}
\end{figure*}

\clearpage
\begin{figure*}[t]
\centering
\includegraphics[width = \textwidth]{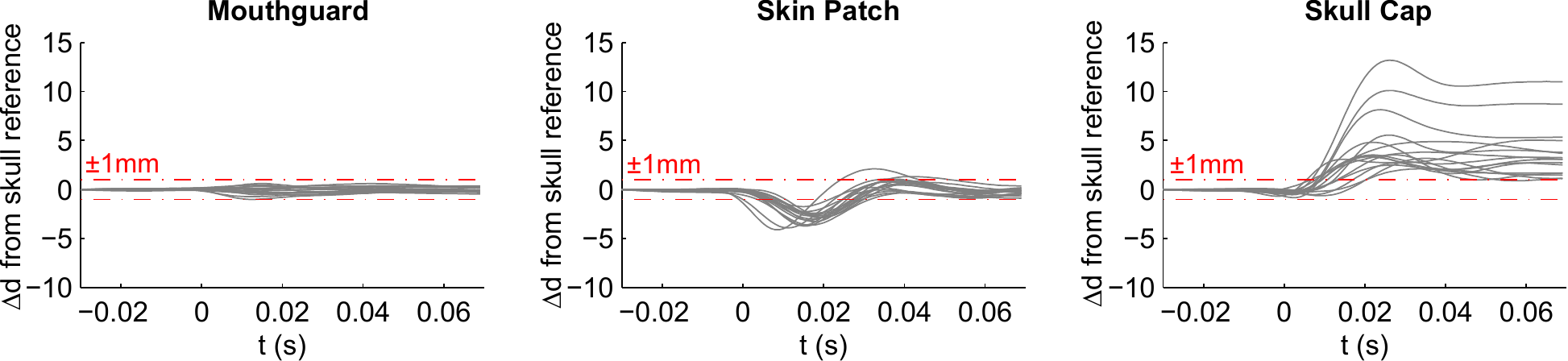}
\caption{
Sensor coupling evaluation. Using high speed video, we compared the relative displacements the three sensors from the skull. Among 16 trials, the mouthguard always had sub-millimeter displacements from the skull within video error, while the other two sensors had higher displacements.
 }
\label{fig:coupling_results}
\end{figure*}

\clearpage
\begin{figure*}[t] 
\centering 
\includegraphics[width = \textwidth]{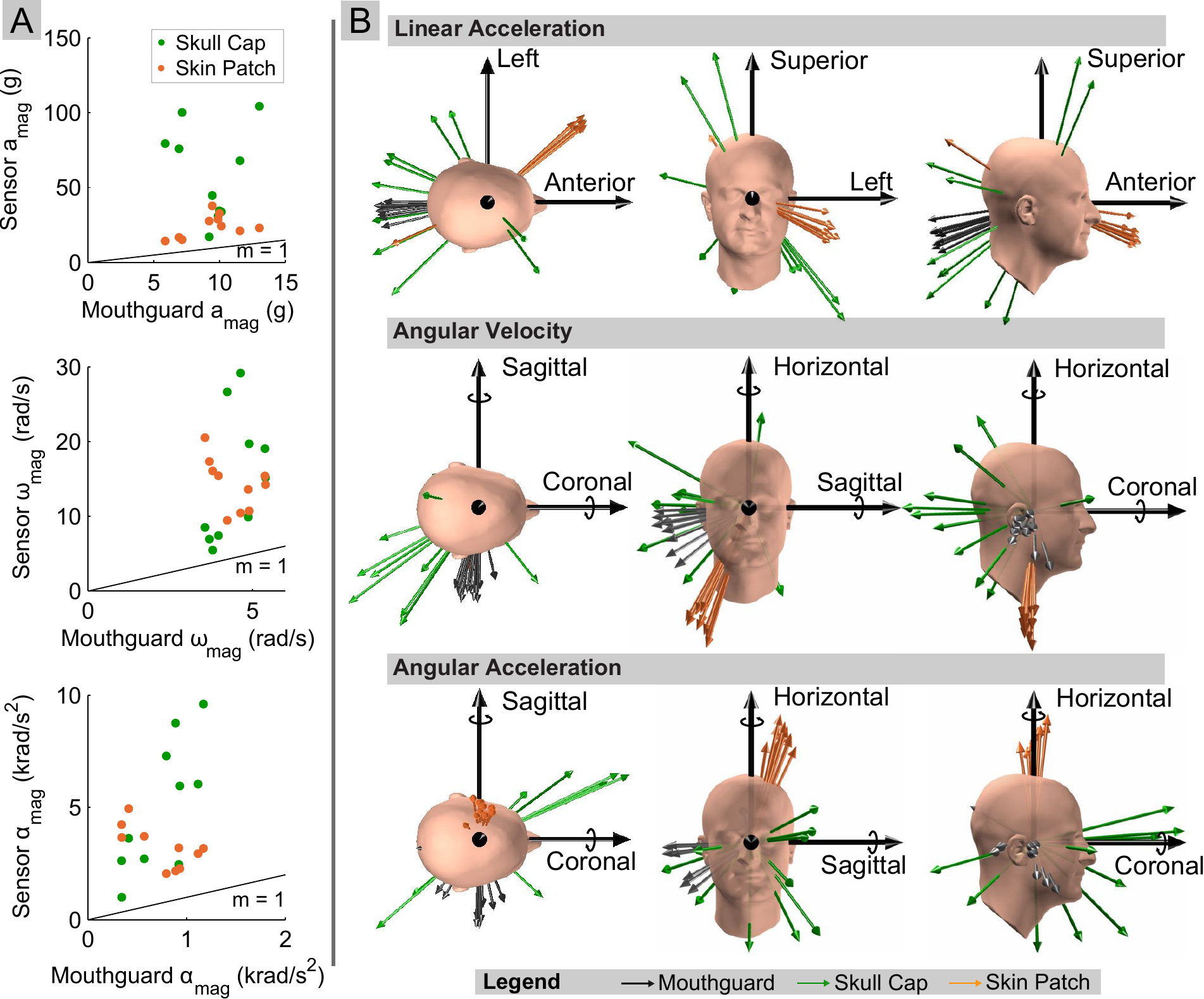}
\caption{Comparing magnitude and direction of peak kinematics. Skin patch and
  skull cap measurements were compared with the mouthguard, which was used as
  the skull reference. (A) shows scatter plots of peak magnitudes of linear
  acceleration, angular velocity, and angular acceleration. The m=1 reference
  line represents the ideal correlation when skin patch and skull cap match the
  reference. In (B), we plot vectors showing the direction of head accelerations
  and velocities at the moment of peak magnitude. The mouthguard, or skull
  reference, measured head motion to exhibit mainly posterior or superior linear
  acceleration, with sagittal rotation. The other two sensors, however, predict
  different directions of acceleration/velocity.}
\label{fig:lin_regression_of_peaks}
\end{figure*}

\clearpage
\begin{figure*}[t]
\centering
\includegraphics[width = \textwidth]{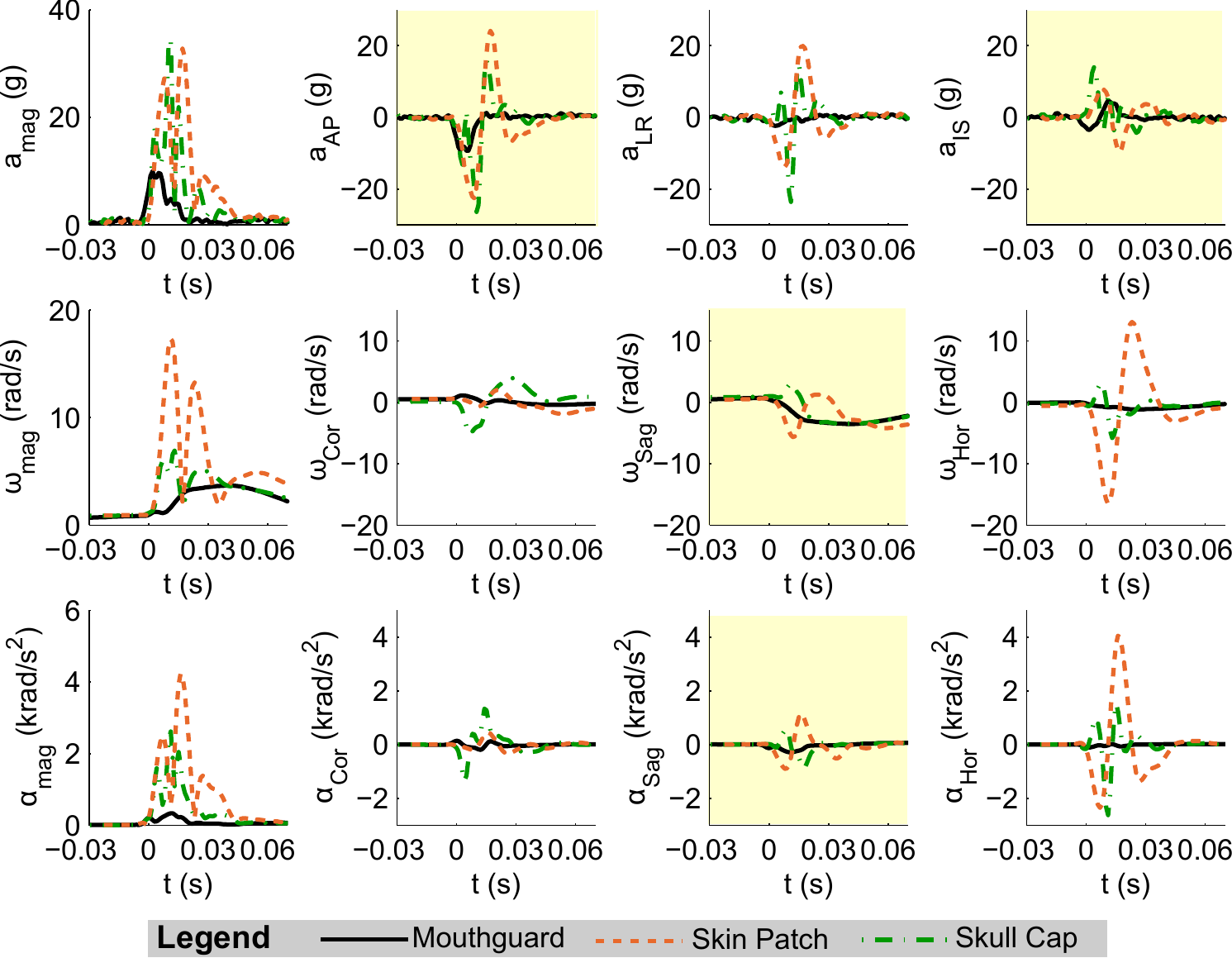}
\caption{
Comparing 6DOF kinematics in a sample impact. Measurements from skin patch and skull cap are compared with the mouthguard reference. Both sensors over-predict accelerations in the sagittal plane (highlighted axes) as well as out-of-plane axes.
 }
\label{fig:accuracy_results_waveforms}
\end{figure*}

\clearpage
\begin{figure}
\centering
\includegraphics[width = 0.8\textwidth]{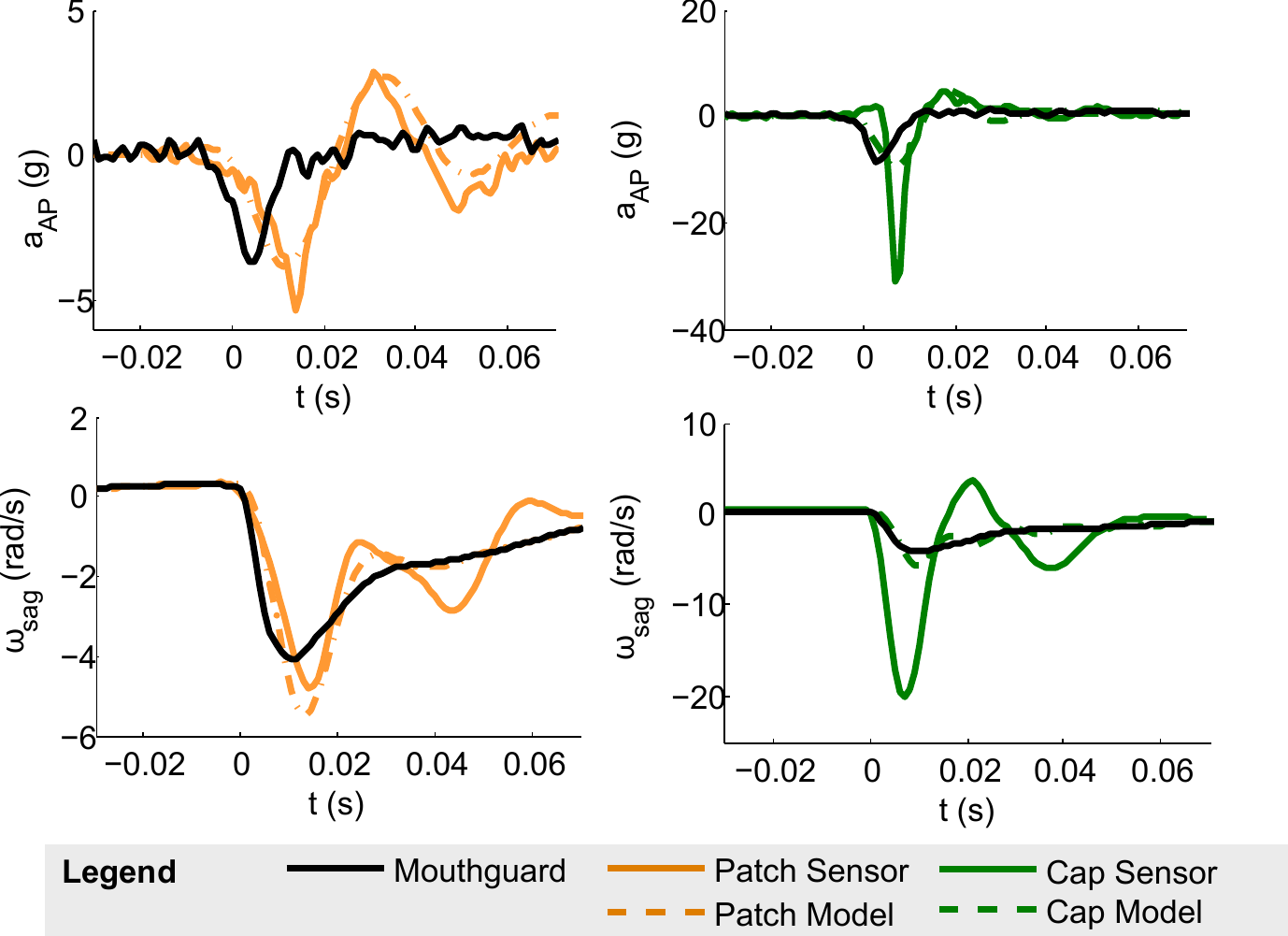}
\caption{Model predictions for skin patch and skull cap. For skin patch AP translation, sagittal rotation, and skull cap AP translation, we could fit underdamped second order linear systems to model sensor output. For skull cap y rotation, the mouthguard (skull) input lagged behind the sensor output, and thus this axis could not be modeled using this simple model.
 }
\label{fig:model_evaluation_waveforms}
\end{figure}

\clearpage
\begin{figure}
\centering
\includegraphics[width = 0.8\textwidth]{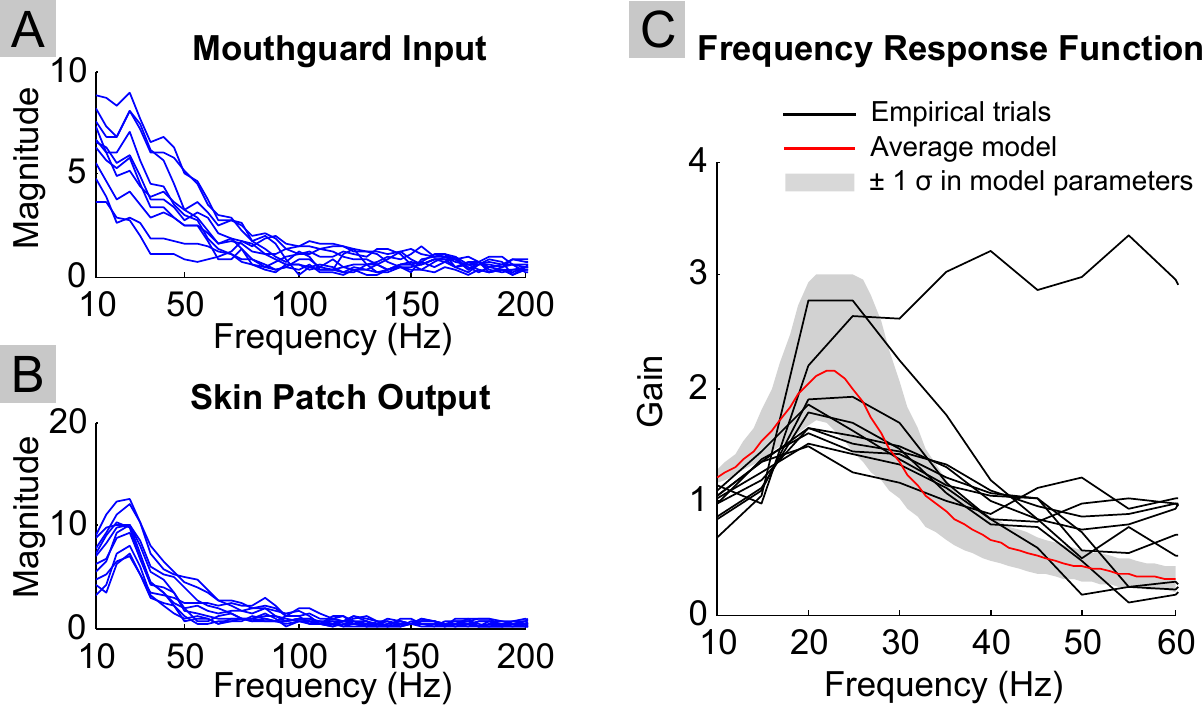}
\caption{Frequency response function of skin patch in AP linear acceleration. The FFT of the mouthguard input (A) and skin patch output (B) both show peak amplitudes occurring in a low frequency range. The frequency response function of the system (C) shows that for 9 of 10 of the trials modeled, the frequency response functions are similar, within 1 standard deviation (shaded region) of the theoretical model (red line), which further demonstrates linearity of the system.
 }
\label{fig:freq_response_fn}
\end{figure}

\setcounter{table}{0}
        \renewcommand{\thetable}{S\arabic{table}}%
\setcounter{figure}{0}
        \renewcommand{\thefigure}{S\arabic{figure}}%

\clearpage
\begin{table*}[ht] 
\caption{Linear Regression of Peak Kinematics}
\centering
\includegraphics[width = \textwidth]{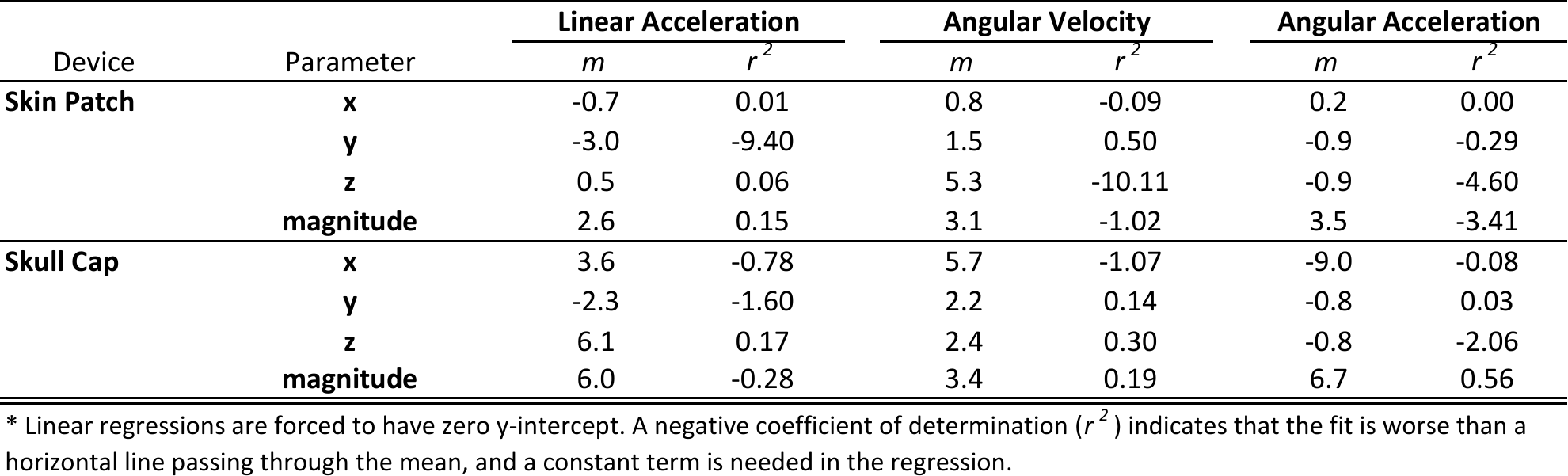}
\label{suptable:lin_reg_results} 
\end{table*}

\clearpage
\begin{table*}[ht] 
\caption{Sensor RMS differences with respect to mouthguard time traces at sensor location}
\centering
\includegraphics[width = \textwidth]{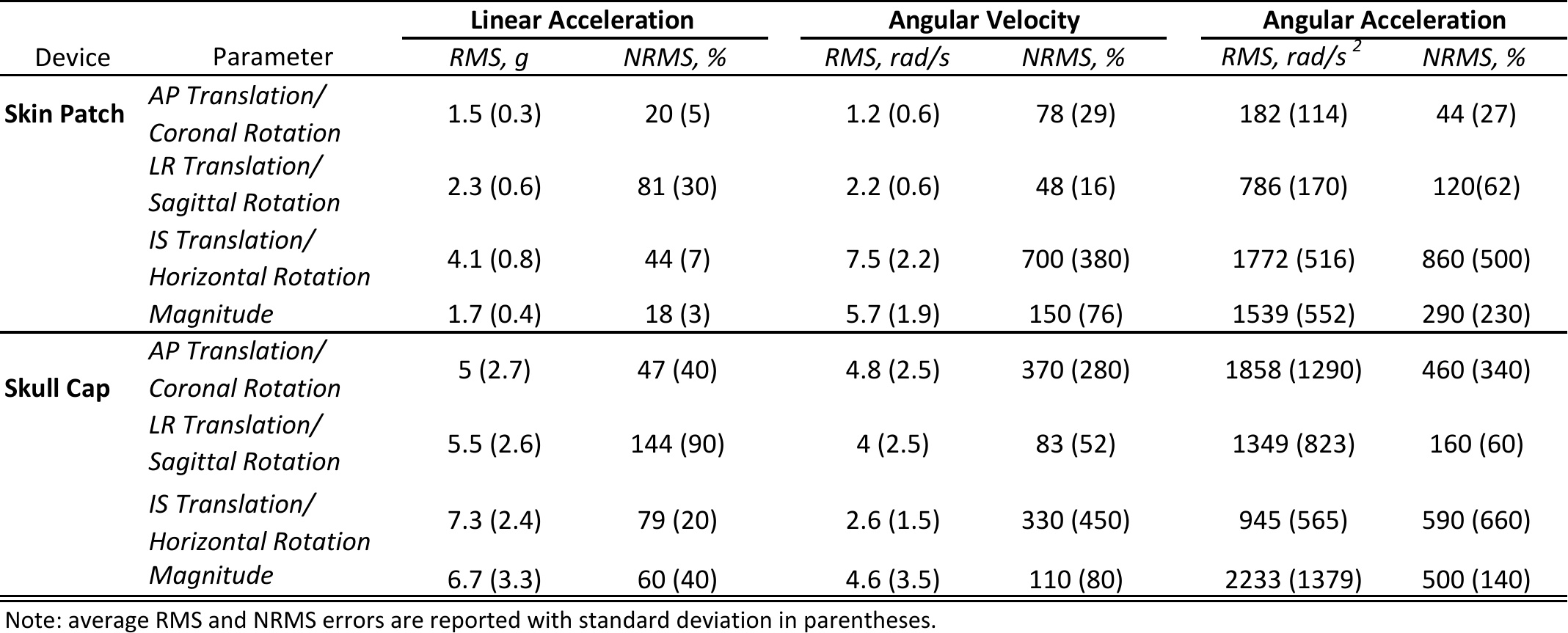}
\label{suptable:RMS_results_at_sensor} 
\end{table*}

\clearpage
\begin{figure}
\centering
\includegraphics[width = \textwidth]{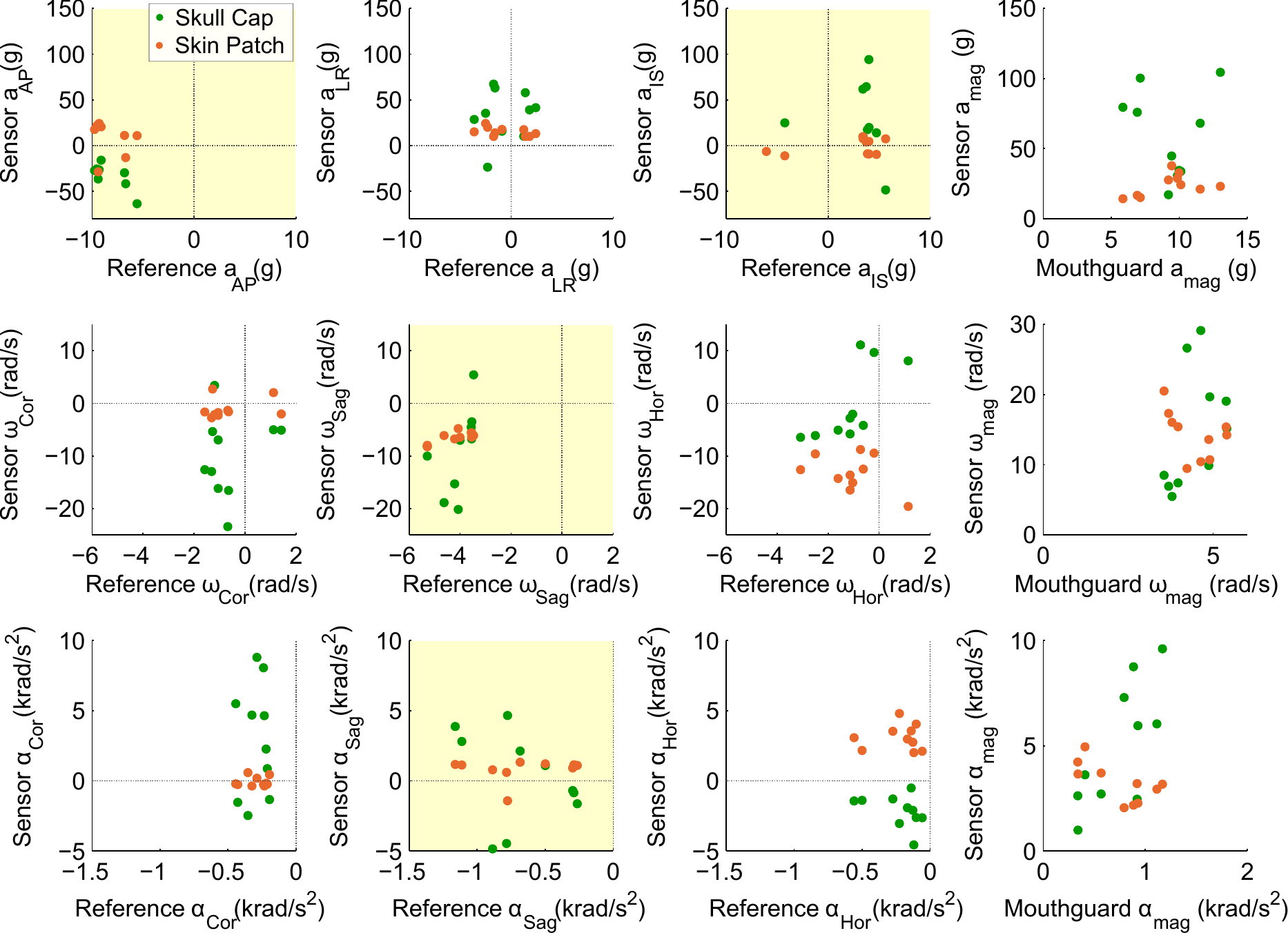}
\caption{Sensor peak measurements projected to head CG. Here we show the peak skin patch/skull cap measurements in each individual degrees of freedom with respect to the mouthguard reference, in addition to the magnitude peaks. Note that for individual axes, the sign of the peak is taken into account. Peaks in the 1st or 3rd quadrants indicate agreement in sign between sensor and reference, while peaks in the 2nd or 4th quadrants indicate opposite signs between sensor and reference.
 }
\label{supfig:peak_scatter}
\end{figure}





%
%
%
\end{document}